\documentclass[11pt,a4paper]{article}
\usepackage[utf8]{inputenc}
\usepackage{lmodern}
\usepackage[onehalfspacing]{setspace}
\usepackage{amsmath}
\usepackage[scale=0.75]{geometry}
\usepackage{graphicx}
\usepackage{color}
\usepackage[small,bf, justification=justified]{caption}
\usepackage{wrapfig}
\usepackage[]{placeins}
\usepackage{hyperref}
\usepackage[backend=biber, style=numeric]{biblatex}
\usepackage{amssymb}
\usepackage{bbm}
\usepackage[gen]{eurosym}
\usepackage{tikz}
\usetikzlibrary{automata,positioning}
\usepackage{subfig}
\usepackage[T1]{fontenc}
\usepackage{algorithm,algpseudocode}
\usepackage[toc,page]{appendix}
\usepackage{caption}
\usepackage{cleveref}
\usepackage{float}
\usepackage{ulem}

\bibliography{Literatur}
\title{Identifying Dominant Industrial Sectors in Market States of the S\&P 500 Financial Data}

\author{Tobias Wand$^{1,2,*}$ \and Martin Heßler$^{1,2}$ \and Oliver Kamps$^{2}$}
\date{%
   \small $^1$Westfälische Wilhelms-Universität Münster, Institut für Theoretische Physik, Germany\\%
    $^2$Westfälische Wilhelms-Universität Münster, Center for Nonlinear Science, Germany\\%
    $^*$Corresponding author: \textit{t\_wand01@wwu.de}\\[2ex]%
   \normalsize August 2022, Revised March 2023
}

\begin{document}

\maketitle

\begin{abstract}
Understanding and forecasting changing market conditions in complex economic systems like the financial market is of great importance to various stakeholders such as financial institutions and regulatory agencies. Based on the finding that the dynamics of sector correlation matrices of the S\&P 500 stock market can be described by a sequence of distinct states via a clustering algorithm, we try to identify the industrial sectors dominating the correlation structure of each state. For this purpose, we use a method from Explainable Artificial Intelligence (XAI) on daily S\&P 500 stock market data from 1992 to 2012 to assign relevance scores to every feature of each data point. To compare the significance of the features for the entire data set we develop an aggregation procedure and apply a Bayesian change point analysis to identify the most significant sector correlations. We show that the correlation matrix of each state is dominated only by a few sector correlations. Especially the energy and IT sector are identified as key factors in determining the state of the economy. Additionally we show that a reduced surrogate model, using only the eight sector correlations with the highest XAI-relevance, can replicate $90\%$ of the cluster assignments. In general our findings imply an additional dimension reduction of the dynamics of the financial market.
\end{abstract}

\paragraph{Keywords:}
Econophysics, Finance,  Explainable AI, Clustering, Market States

\section{Introduction}

The economy is an example of a complex system with a large number of  interacting parts \cite{MantegnaStanleyBook}. Its non-stationarity makes the analysis much more difficult than for a stationary system. Nevertheless the time evolution of many non-stationary complex systems can be described as a sequence of transient states, i.e. the system passes through a sequence of states where it remains for a short time, before transitioning to the next state. Since  this represents an enormous reduction of the dimensionality of the system, the identification of these states might open the possibility  for a better prediction and control of the system. Examples can be found in such diverse fields as e.g. neurodynamics \cite{Ashwin2005nat} and fluid dynamics \cite{Fernex2021}. Since in many cases an underlying mathematical description is not available, one has to rely on data driven methods to find such states, e.g. via clustering or similar methods \cite{Hutt, GrabenHutt_PRL}. Due to the increasing availability of data for the economy nowadays, such methods can also be applied in this field of research. In \cite{Mnnix2012} the authors tried to identify such states in the economy by analysing correlation matrices of daily S\&P 500 data. Generally, the correlation matrix between the returns of the different stocks is an object of extensive research \cite{BROWN1989, RandomMatrixLalouxPotters,MantegnaCorrelations,Markowitz1952,Review_CorrelationsHierarchiesClustering_FinancialMarkets,RandomMatrixStanley}. \\
In \cite{Mnnix2012,Rinn2015DynamicsOQ,Stepanov2015} the correlation matrices of the daily S\&P 500 data could be sorted into eight clusters, which can be interpreted as states of the economy corresponding to e.g. economic growth phases or crises \cite{Mnnix2012} and show locally stable behaviour \cite{Rinn2015DynamicsOQ,Stepanov2015}. Further research showed that subtracting the dominant market mode identified in \cite{RandomMatrixLalouxPotters,RandomMatrixStanley} from the correlation matrix leads to a reduced-rank correlation matrix that highlights the differences between different market states and allows to identify exogenous events, precursors for crises and collectivity of industrial sectors \cite{Heckens_2020,Heckens_2022,Heckens_2022_Arxiv}. An alternative approach to the identification of market states is described in \cite{Marsili_FinancialStatesViaReturns}, where the author uses the clustering algorithm given in \cite{Marsili_alg}. While this algorithm uses the correlation matrices between the observations that would be sorted into a proposed clustering structure $s$ to infer the optimal parameters of the clustering algorithm, the observables of interest are the returns of the individual companies. Note that an elegant interpretation of the individual market states is given in \cite{Marsili_FinancialStatesViaReturns}. The states are interpreted regarding the mean returns of the companies conditional on the trading day being in a specific state $s$. Thus, among other things, the author finds that the Oil \& Gas and Petroleum sectors have a dominating influence for one of the temporal clusters.\\
The correlation matrices are also the focus of observation in \cite{Mnnix2012} where eight market states are identified. In \cite{Mnnix2012} the differences between the market states are qualitatively discussed by comparing their centroids to each other and found different regimes of correlation structure: some states correspond to strong overall correlations throughout the market, whereas other states exhibit weak correlations outside of companies from the same industry sector or anticorrelations between some sectors. However, these comparisons were purely qualitative via plotting and comparing heat maps of the clusters' centroids and thus, there has not yet been a quantitative analysis of what differentiates these market states from each other. In this article our goal is to identify the dominant factors within the different economic states in the framework of \cite{Mnnix2012,Rinn2015DynamicsOQ,Stepanov2015} by using a method \cite{Kauffmann2022} from the emerging field of Explainable Artificial Intelligence (XAI) which can be understood as a collection of methods that shed light on the black box that most AI methods seem to be from the user's perspective \cite{molnar2022,XAI_book}. XAI allows a quantitative ranking of how much a given observable has influenced the decision algorithm (here: clustering algorithm) and therefore gives us deeper insights into the differences between the market states than the analysis in \cite{Mnnix2012}. Most importantly, this methodology can allow us to perform a dimensionality reduction by identifying the most important sector correlations without any qualitative ambiguity. XAI is nowadays considered for applications in physics \cite{XAI_MaterialScience,XAI_HEP,XAI_ParticlePhysics} and finance alike \cite{WhitepaperXAIFinance} and in particular, financial regulators have already embraced the methods of XAI to increase transparency for customers with regards to AI-supported business decisions of financial institutions \cite{Bafin}. 

In \cite{Marsili_FinancialStatesViaReturns} the market states are characterized and interpreted in terms of the clustered returns, whereas we consider the correlations of different market sectors. Such an analysis is similar to the qualitative analysis of the states in \cite{Mnnix2012,Rinn2015DynamicsOQ,Stepanov2015}, but XAI can give us a more detailed analysis: The observables in \cite{Mnnix2012,Rinn2015DynamicsOQ,Stepanov2015} and in our study are the correlation matrices and therefore, the clustered variables are the correlations between different sectors. These correlations, however, are strongly interdependent; not only because of the economic real-world situation that they model, but also because of mathematical constraints of the correlation matrix (its eigenvalues are non-negative, meaning that its matrix entries cannot be chosen independently from each other; cf. e.g. \cite{CorrelationMatrices} for the problems that can arise from this fact in practical applications). The XAI algorithm from \cite{Kauffmann2022} is well-suited to deal with such dependencies because of its holistic approach: it starts with the prediction of the k-means algorithm and goes backwards through the algorithmic structure until it arrives at the input layer, where it distributes the prediction relevance on all input variables (here: correlations) simultaneously. I.e. it does not look at an isolated input variable in a vacuum, but always takes the values of other (potentially highly interdependent) variables into account.

Here, based on the hypothesis that the dynamics of the financial market can be described by passing through a sequence of a limited number of transient states, we use the k-means clustering method to identify the clusters. While most of the above clustering analyses use a hierarchical k-means clustering, we use regular k-means clustering. This has a big advantage because it can be reformulated to allow the use of XAI, which can analyse the decisions of the clustering algorithm \cite{Kauffmann2022}. This allows us to reveal the dominant factors of the different market states and to compare the market states quantitatively, thereby improving the qualitative comparison depicted in figure 2 of \cite{Mnnix2012}. Moreover, as shown later in the article, the XAI selection allows for a dimensionality reduction of the system by selecting the most relevant correlations.

The remainder of this article is structured as follows: Section \ref{sec:DataPrep} explains how the S\&P 500 data was collected and preprocessed so that it could be used for our analysis. Section \ref{scc:MainPart} introduces the clustering method and the XAI method before interpreting the XAI results. Also, it discusses a reduced model which only uses the features (i.e. sector-sector correlations) with the best XAI values to learn the clustering classification to test the usefulness of XAI analyses for reducing the dimensionality of the data. Finally, section \ref{sec:Discussion} summarises the results, compares them to previous works and suggests follow-up research. The appendix includes some additional information on the XAI algorithm in appendix \ref{sc:APP_LRP}, some additional depictions of results in further figures in appendix \ref{sc:App} and technical details on neural networks in appendix \ref{sc:App_NN}.

\section{Data Preparation}
\label{sec:DataPrep}
The S\&P 500 is a stock index containing 500 large US companies traded at American stock exchanges. Daily stock data from these companies were downloaded via the Python package \textit{yfinance} \cite{yfinance} for the time period between 1992 and 2012, which is the same time period as in \cite{Rinn2015DynamicsOQ}. Only stock data of companies that were part of the S\&P 500 index during at least 99.5\% of the time were used for this analysis. Following the approach in \cite{Stepanov2015}, the data is aggregated according to the 10 GICS (Global Industry Classification Standard) sectors representing individual parts of the economy \cite{GICS}. Stock prices from companies of the same GICS sector were added to calculate representative time series $(S_t)_{t=1,\dots,T}$ for each of the 10 sectors by adding the price time series of the companies in the same economic sector, which corresponds to the price of buying one stock of each. This drastically reduces the number of free parameters $N(N-1)/2$ in the $N\times N$ correlation matrix from an order of $10^4$ to exactly $45$ to ensure that there is enough data to estimate each correlation, i.e. that the estimation is not underdetermined. If a company's price time series is not available for the full time period, the remaining 0.5\% of the data are interpolated linearly with the \textit{.interpolate()} method in \textit{pandas} \cite{reback2020pandas, mckinney-proc-scipy-2010}. Overall, there are 249 companies' time series for 5291 trading days. Each sector's returns $R_t = (S_{t+1} - S_t)/S_t$ are locally normalised to remedy the impact of sudden changes in the drift of the time series with the method introduced in \cite{SCHAFER20103856} as
\begin{equation}
\label{eq:normalisation}
    r_t = \frac{R_t - \langle R_t\rangle_n}{\sqrt{\langle R^2_t\rangle_n - \langle R_t\rangle^2_n}}.
\end{equation}
Here, $\langle\dots\rangle_n$ denotes a local mean across the $n$ most recent data points, i.e. $r_t$ is subjected to a standard normalisation transformation with respect to the local mean and standard deviation (i.e. volatility $\sigma$). Following \cite{Mnnix2012}, $n=13$ was chosen for the daily data. 
For each time step $t$ and each pair of sectors $i,j$ the local correlation coefficients 
\begin{equation}
\label{eq:Correlation}
    C_{i,j} = \frac{\langle r^{(i)}_t r^{(j)}_t\rangle_\tau - \langle r^{(i)}_t \rangle_\tau  \langle r^{(j)}_t \rangle_\tau}{ \sigma^{(i)}_\tau \sigma^{(j)}_\tau  }
\end{equation}
are calculated over a time period of $\tau=40$ trading days like in \cite{Mnnix2012} (the 40 working days correspond to two trading months) with the local standard deviations $\sigma^{(i)}_\tau$. The window is shifted by one trading day each, resulting in overlapping windows, and hence, the window to calculate $\langle\dots\rangle_n$ in equation \ref{eq:normalisation} is also shifted by one day. This approach is similar to the sector analysis in \cite{Rinn2015DynamicsOQ}, where the correlations between the individual companies' stock time series are averaged within their GICS sectors. However, our procedure of first computing the GICS sector time series and second, calculating the correlation ensures that larger companies are given more weight in the resulting correlation matrix than smaller ones.\\
It has to be stressed that the GICS sector classification has changed over the considered time period. E.g. when \cite{Rinn2015DynamicsOQ} was published, there were only 10 GICS sectors, whereas nowadays, there are 11. During the data period from 1992 to 2012, the exact composition of sectors and subsectors has also changed which poses additional problems in sorting a company into a sector. Here, the historical data is sorted into the sectors according to the companies' GICS classification at the end of the time period in 2012, reflecting the same sectors as the ones used in \cite{Mnnix2012}.

\section{Identifying the most relevant sector correlations within a market state}
\label{scc:MainPart}

In this section we first give an overview of the clustering method with further technical details explained in the appendix \ref{sc:APP_LRP}, before using XAI to shed light on how the clustering algorithm arrives at its results. Then, the XAI results are discussed and interpreted. Finally, a reduced surrogate model that only uses the highest-XAI features is used to see whether the reduction to features with high XAI preserves most of the clustering algorithm's information.

\subsection{Identifying Market States via K-Means Clustering}

Because the analyses \cite{Mnnix2012, Rinn2015DynamicsOQ, Stepanov2015} unanimously revealed eight market states for the time period analysed in this article, we set the number of clusters for the k-means algorithm to eight. In contrast to the procedure in the literature, however, regular k-means clustering is used instead of a hierarchical binary k-means clustering in order to make the cluster centres amenable to the XAI method suggested in \cite{Kauffmann2022}. The regular k-means clustering method starts with a predefined number of $k$ clusters and randomly selects $k$ initial cluster centres before iteratively assigning each daily correlation matrix to its nearest cluster centre and updating the centres by setting them as the cluster's centroid until their centroid positions converge \cite{KMeans}. The market correlation matrix $\textbf{C}$ belongs to cluster $j$, if for the cluster centroids $(\textbf{C}_l)_{l=1,\dots,k}$ the inequality
\begin{equation}
\label{eq:KMeansInequality}
    ||\textbf{C}-\textbf{C}_j|| < \underset{l\neq j}{\min} ||\textbf{C} - \textbf{C}_l||
\end{equation}
holds for a suitable distance metric $||\dots||$. Here, the Euclidean distance is used.

\begin{figure}
    \centering
    \includegraphics[width=0.8\textwidth]{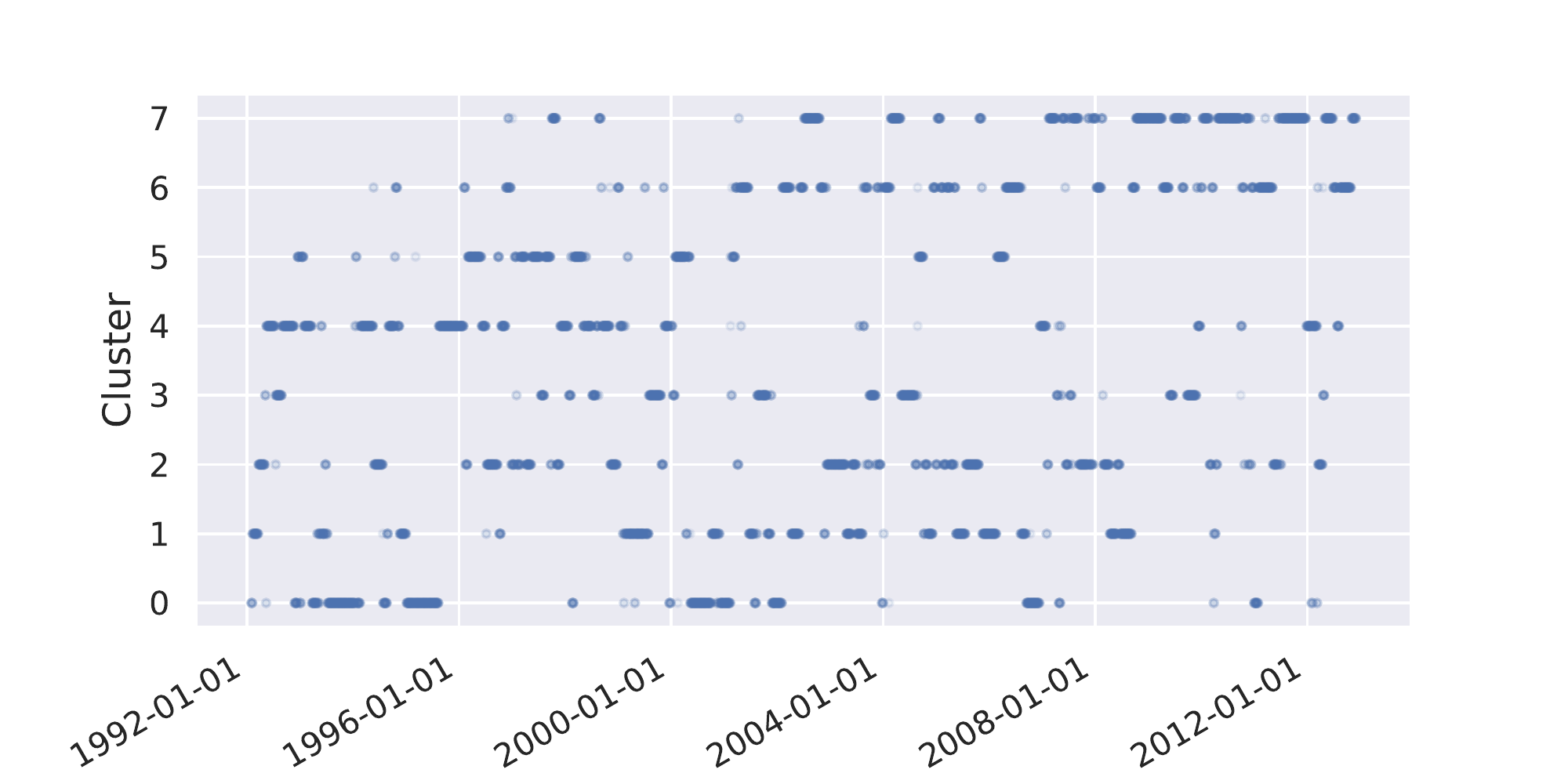}
    \caption{This figure shows which cluster state is occupied by the market at which point in time. Individual data points are depicted with low opacity and the slight overlap between adjacent points increases the opacity and may appear like a solid line.}
    \label{fig:Cluster_vs_time}
\end{figure}

The time series of the resulting clusters is depicted in figure \ref{fig:Cluster_vs_time} and shows a qualitatively similar behaviour to those in \cite{Mnnix2012, Rinn2015DynamicsOQ,Stepanov2015}. In the observed time period, some clusters cease to appear more than sporadically in the data after some years (clusters 0, 1 and 5) and some only start to appear more than sporadically years after the start of the observation time period (cluster 6, 7 and again 5). However, figure \ref{fig:Cluster_vs_time} shows some quantitative differences compared to the aforementioned literature, in particular because the clusters in figure \ref{fig:Cluster_vs_time} often appear isolated and for short time periods apart from their main emergence interval. This may be attributed to the slightly different clustering technique used for this article and the three clustering analyses \cite{Mnnix2012, Rinn2015DynamicsOQ,Stepanov2015} also show differences in their plots of clusters against time.

\subsection{Explainable AI for Clustering}
\label{sec:Results_XAI}

Given the results for the clusters we now want to explore whether the clusters are dominated by certain industrial sectors and, if this is the case, which sectors are significant for the identification of the market states represented by the clusters. To this end we use the method developed in \cite{Kauffmann2022} to analyse k-means through the lens of XAI. Therefore two steps are necessary \cite{Kauffmann2022}: First, the clustering classifier has to be neuralised, i.e. rewritten as a neural network. Then, an XAI method for neural networks is used for analysing the neuralised clustering algorithm, namely the Layer-wise Relevance Propagation (LRP) \cite{LRP_book}. For each instance (data point), the XAI algorithm estimates how much the values of its features contributed to the neural network's decision for this instance. It answers the question which particular feature values led to $\textbf{C}(t)$ being sorted into cluster $j$ instead of any other cluster. The contribution of the $i^\textmd{th}$ feature is quantified in a relevance score $\rho_i$ which is calculated according to equations (2) and (4) from \cite{Kauffmann2022} for each instance with respect to its cluster assignment $j$ for all features $i$. Appendix \ref{sc:APP_LRP} gives further details on the implementation of this algorithm.


\subsubsection{XAI: From Local to Global Relevances}
\label{sec:RelevantCorr}
\begin{figure}
    \centering
    \includegraphics[width =  \textwidth]{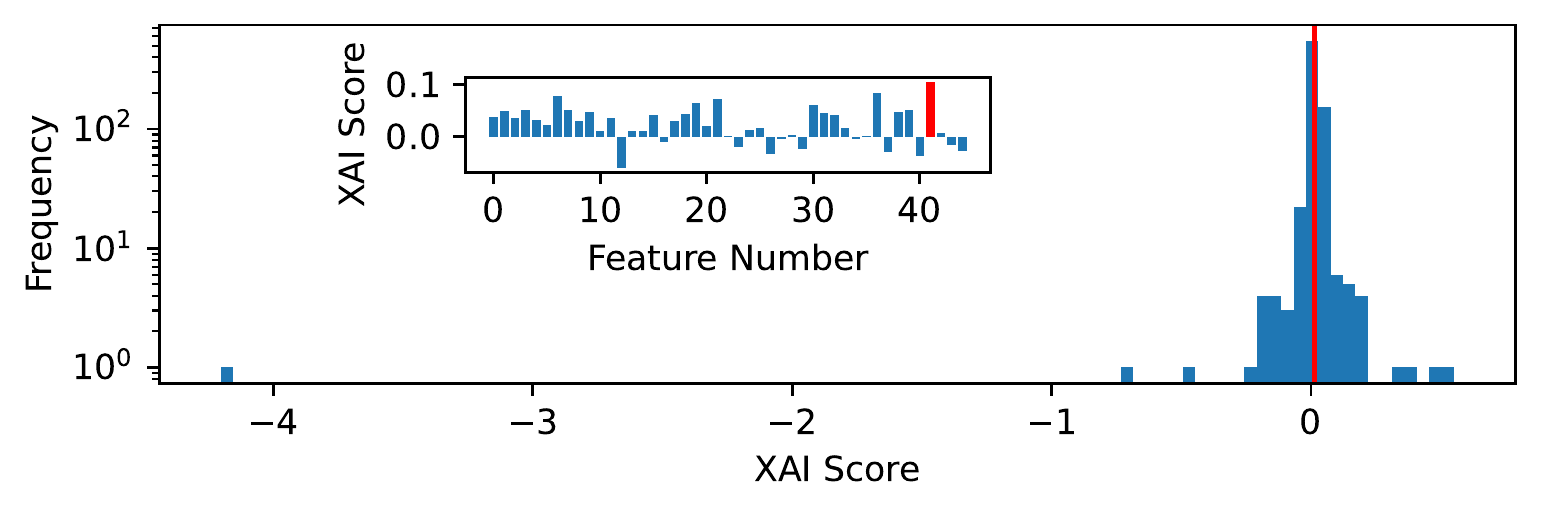}
    \caption{Inset: Barplot of the XAI importance scores for the the first day in cluster 2, showing how much each feature contributed to the classification. The mode (feature with highest XAI score) is depicted in red. Main figure: Histogram of the XAI importance of the correlation Energy/Materials for all instances (trading days) in cluster 2. Outliers like the one visible on the far left of this histogram distort the mean. The median value is depicted in red.}
    \label{fig:HistogramXAI}
\end{figure}

This method produces relevance scores $\rho_i$ for each feature $i$ (for the correlation between sectors) for every instance (i.e. for each trading day), as depicted in the inset of figure \ref{fig:HistogramXAI} for the first day of cluster 2. However, we wish to focus on the aggregated relevance of one specific feature to understand which sector correlations are more important for the market state than others. As an example, figure \ref{fig:HistogramXAI} shows the histogram of XAI relevances for the correlation Energy/Materials for all trading days in cluster 2. Are such values particularly high? Does the correlation between Energy and Materials matter more for cluster 2 than other correlations? To answer such a question, one has to use the XAI values of each instance (local values) for a global aggregation to compare the aggregated influence of the Energy/Materials correlation with other correlations.
Using the mean relevances of each sector seems reasonable, but the mean is distorted by some instances with very high absolute values for some of their relevances. Hence, one can instead focus on the median of each feature's relevance for each cluster's instances. Alternatively, one can identify each instance's feature with the highest $\rho_i$ and then concentrate on the features that are most frequently the most important feature of an instance within each cluster. Because it regards the statistical mode (i.e. the most frequently observed value) and calculates the mode of the mode of each instance's $\rho_i$, we chose to name this method "mode-mode". The next sections addresses the question, which of these aggregation methods is more suitable for the data.

\subsubsection{Determining Relevant Correlations}

\begin{figure}
    \centering
    \includegraphics[width = \textwidth]{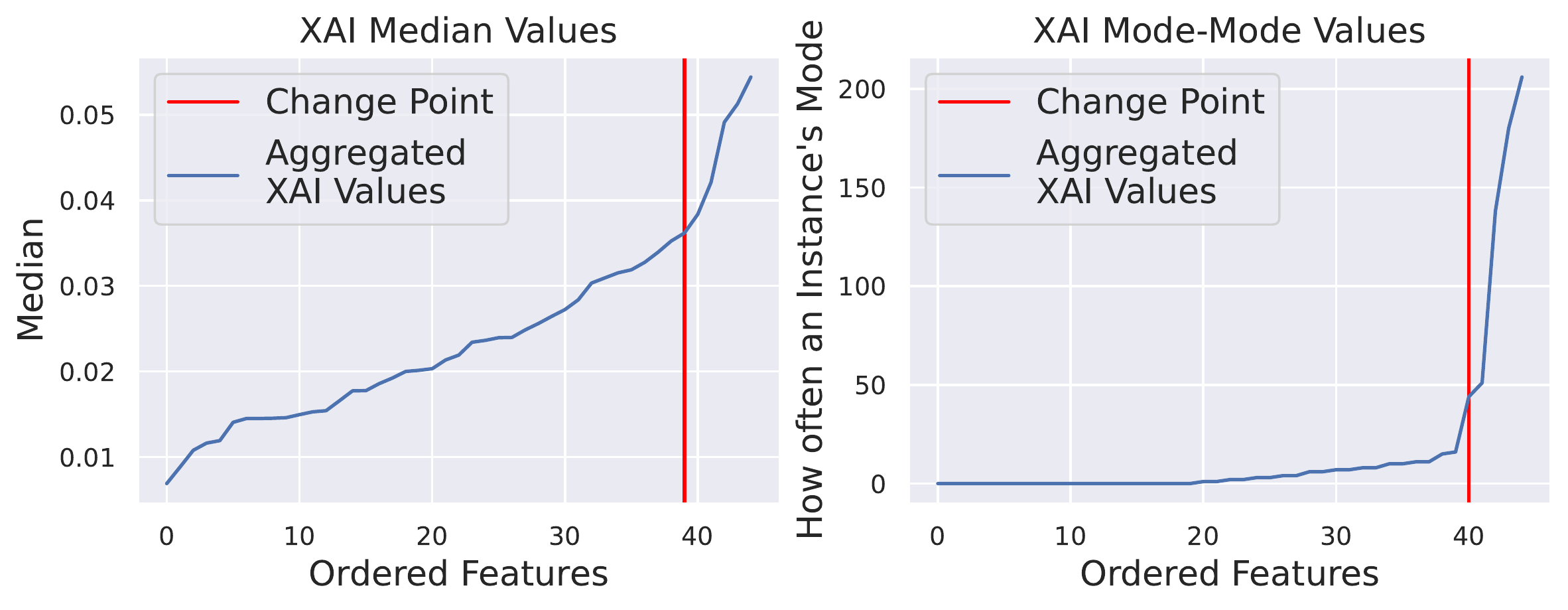}
    \caption{Comparison of ascendingly plotting the features according to the median values XAI (left) and mode-mode XAI (right) in blue and their respective change point estimation with \textit{antiCPy} in red. Plots for the other clusters are included as supplementary material in the appendix \ref{sc:App}.}
    \label{fig:ElbowPlots_Mean_ModeMode}
\end{figure}

For each individual cluster, the XAI relevances of each instance are aggregated to analyse the importance of any feature for the clustering decision via two methods: Either the median $\hat{\rho}_i$ of the $i^\textmd{th}$ feature or the mode-mode method is used. The latter method calculates how often $n_i$ the $i^\textmd{th}$ feature is the most relevant feature of an instance and ranks the features according to their value $n_i$. Both the $(\hat{\rho}_i)_i$ and $(n_i)_i$ values are sorted ascendingly as shown in a sample plot in figure \ref{fig:ElbowPlots_Mean_ModeMode}. These plots can be approximated by piecewise linear segments and sometimes, they show a characteristic elbow shape: They start at a long plateau of low values and later experience a drastic increase, resulting in an almost rectangular shape (cf. figure \ref{fig:ElbowPlots_Mean_ModeMode}, right). This shape allows to divide the features into two groups. Therefore, a change point analysis can identify the most drastic change in the slope of those plots and help us to achieve a clear threshold between important features with high $\hat{\rho}_i$ or $n_i$ and features with little importance \cite{b:linden2014}. Here we use the Bayesian change point analysis implemented in the \textit{antiCPy} package \cite{MartinHeßler} based on the ideas in \cite{a:dose2004}.\\
Like depicted in figure \ref{fig:ElbowPlots_Mean_ModeMode}, the mode-mode values usually show a much more pronounced elbow that makes change point selection easier and their estimated change point is often at a much higher feature number than for the median method. Therefore, if features on the right-hand side of the change point are to be interpreted as highly relevant features, then the mode-mode method allows for a more trustworthy and more exhaustive feature reduction than the median method.

\begin{figure}
    \centering
    \includegraphics[width=\textwidth]{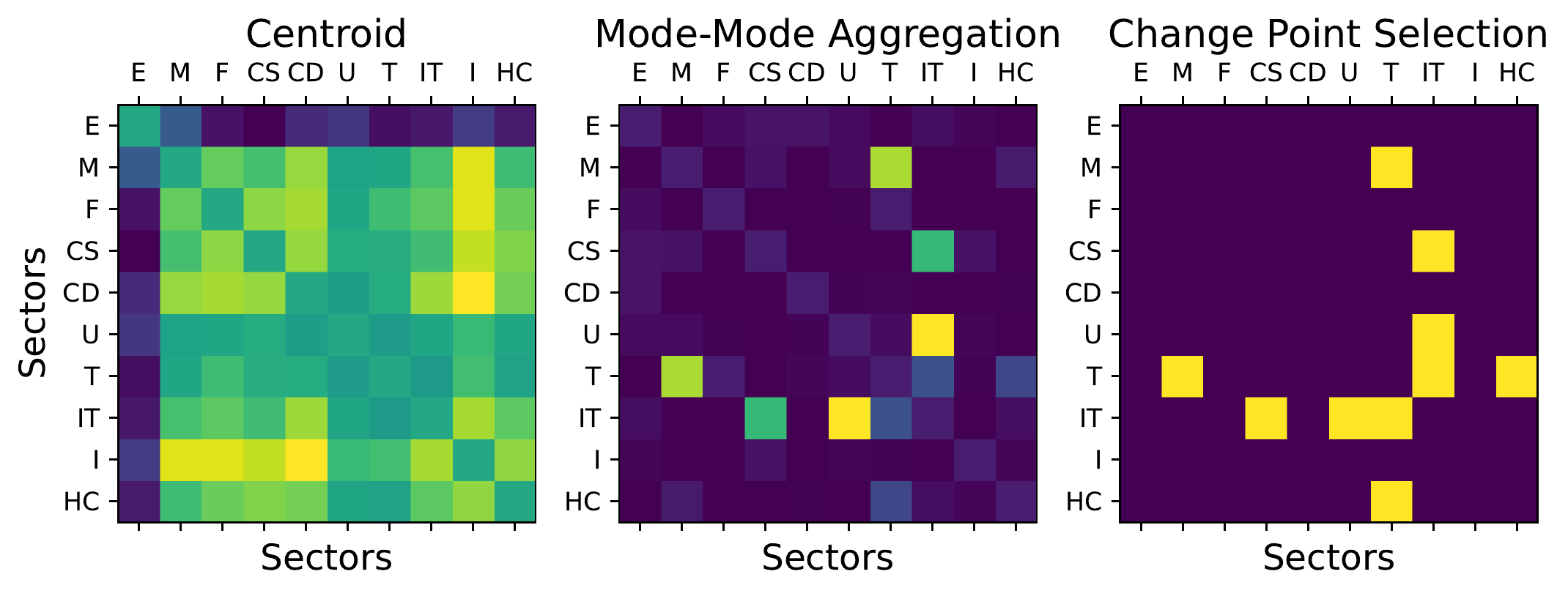}
    \caption{Visualisation of different matrices that represent cluster 2: Comparison between the cluster's centroid, the mode-based aggregated XAI values and the binary antiCPy-selected relevant XAI values. Brighter colours denote higher values. Because the diagonal's correlation values are always 1, their values have been set to the matrices' averages.}
    \label{fig:Compare_Centroids_XAI}
\end{figure}

\begin{figure}
    \centering
    \includegraphics[width =\textwidth]{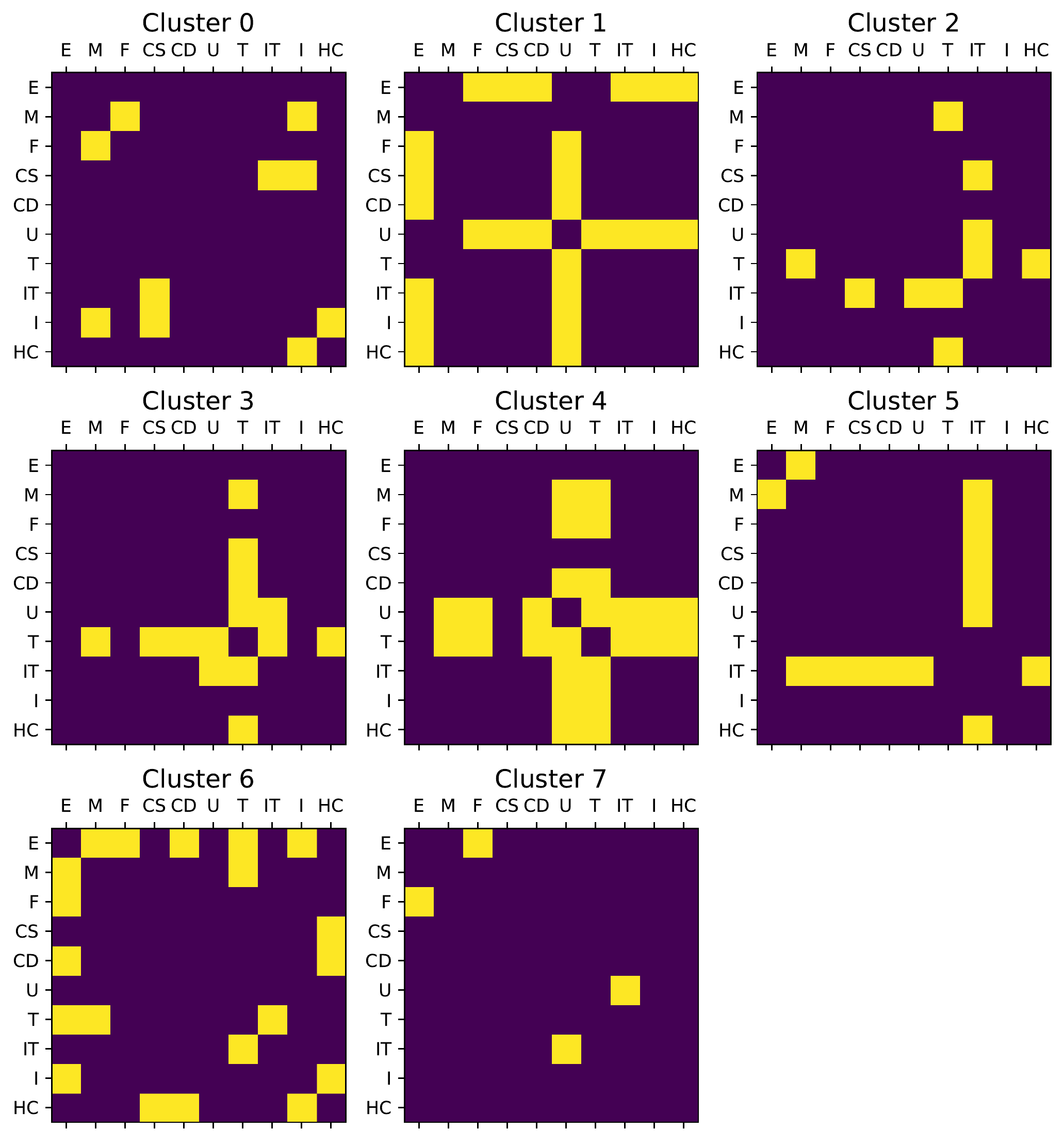} \caption{Relevant correlations (yellow) of the sectors for the 8 clusters in figure \ref{fig:Cluster_vs_time}. Relevance selection was done via the change point analysis explained in section \ref{sec:RelevantCorr} on the mode-mode XAI values. The sectors' abbreviations are
   E: Energy, M: Materials, F: Financial Services, CS: Consumer Staples, CD: Consumer Discretionary, U: Utilities, T: Telecommunication Services, IT: Information Technology, I: Industrials and HC: Health Care. Note that the diagonal has been set to zero relevance by default because its correlations are always 1.}
    \label{fig:RelvantCorrelations}
\end{figure}

We use the mode-mode method to differentiate between relevant and irrelevant correlations, because the approach usually leads to a more parsimonious feature reduction (i.e. higher interpretability) and often much clearer change points (cf. elbow-like relevance plots in \ref{fig:ElbowPlots_Mean_ModeMode}). Note that a relevant correlation is not necessarily unusually strong or weak, but rather particularly useful to differentiate between this cluster and the others. As an example, figure \ref{fig:Compare_Centroids_XAI} shows the centroid (left), the aggregated XAI values (centre) and the change point selection of relevant features (right) for cluster 2. The relevant features cannot be detected trivially by e.g. just identifying particularly high or low values in the centroid, but follow from a comparison to the other centroids. Hence, the change point analysis of relevant values does not follow trivially from the cluster's centroids, but uncovers new information about the data and yields a parsimonious selection of relevant economic sector correlations.\\
A plot of all relevant correlations for each cluster is shown in figure \ref{fig:RelvantCorrelations}. The correlations of the Energy sector appear in states associated with clusters 1 and 6, the Utilities sector in 1 and 4 and the Telecommunication sector in 3 and 4. The IT sector dominates cluster 5 , influences cluster 0 and, alongside Telecommunication Services, also influences cluster 2. States in cluster 7 only have a high relevance for the correlations between Energy/Financial Services and IT/Telecommunication. Other clusters show a high influence of Utilities and Telecommunication Services, which are both tied to the Energy or IT sector, respectively: the Utilities sector includes companies which distribute gas and electricity, while Telecommunication Services are related to the physical infrastructure of information transportation.

\subsection{Dimensionality Reduction for all Clusters}

\begin{figure}
    \centering
    \includegraphics[width =0.99 \textwidth]{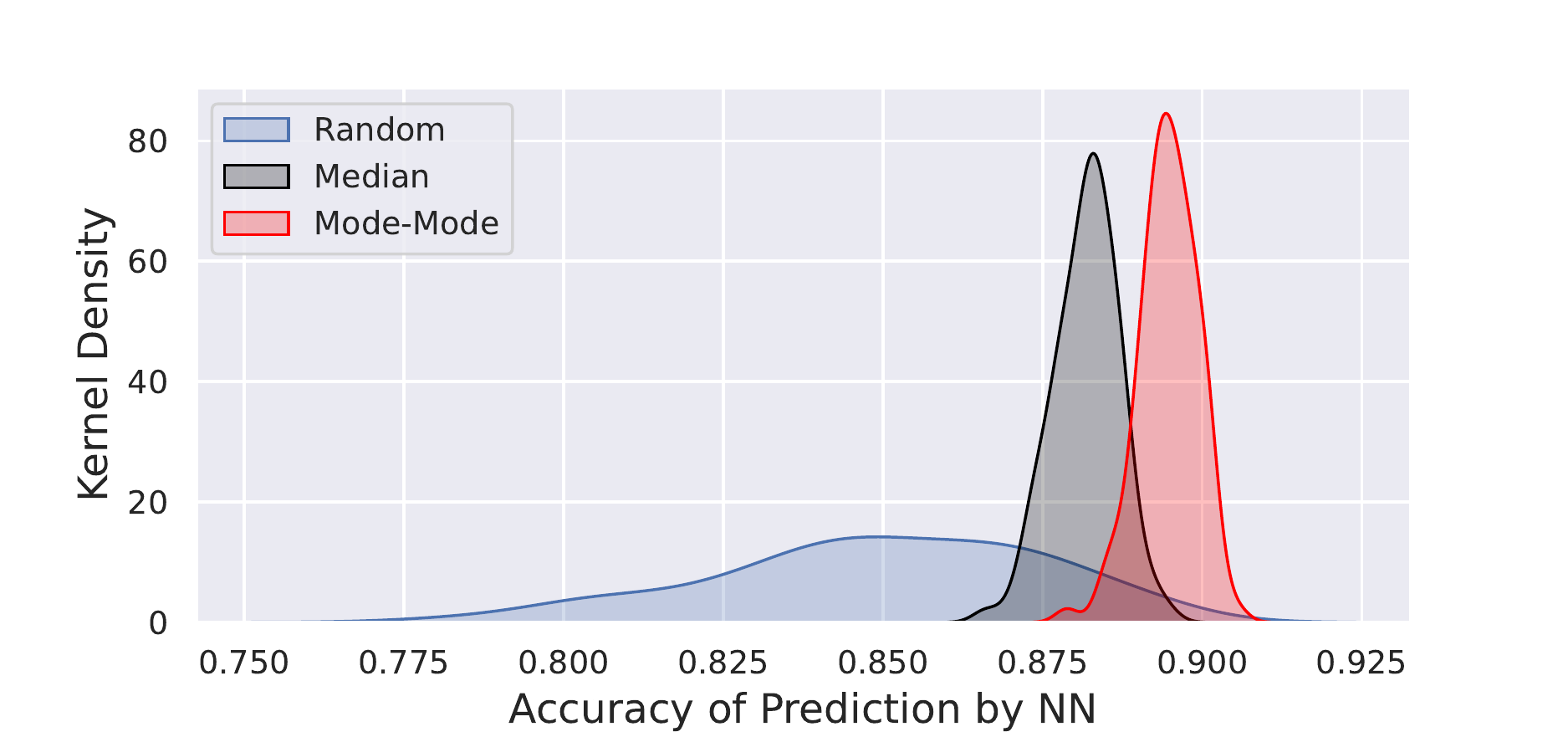}
    \caption{With the eight mode-mode features, the eight best median features and eight random features, 100 neural networks were trained each with a different seed. Kernel density estimations show that the mode-mode and median networks achieve a much higher accuracy than the randomly selected suboptimal features.}
    \label{fig:NN_comparison}
\end{figure}

It makes sense to assume that features with a high relevance should be particularly useful for predicting the original classifier. In particular, this might reveal that only few features are necessary to predict the majority of the cluster assignments. If we only wish to classify one cluster vs. all the others, then the relevant features selected by the cutoff in figure \ref{fig:ElbowPlots_Mean_ModeMode} are a reasonable choice for reducing the dimensionality of the data. But for an all vs. all classifier, choosing all relevant features of all clusters would barely reduce the number of features at all, because clusters 1 and 4 (cf. figures \ref{fig:Appendix1} and \ref{fig:Appendix4}) would already each contribute more than ten features to this list of relevant features. Instead, one can restrict the analysis to the single best feature for each cluster, resulting in eight features instead of the original 45. This also constitutes a second chance to compare the mode-mode selection to the median selection. If a model trained on such a dimensionality reduction predicts most of the cluster assignments correctly, it will serve as a self-consistency check by indicating that features with high XAI relevance are also highly useful for predicting. \\
To test, whether a dimensionality reduction is possible, we train a neural network as a surrogate model that should predict the original clustering decisions, but restrict its data to only some of the original features (cf. \cite[chapter~8.6]{molnar2022}). For each of the eight clusters, only the feature that has the highest XAI relevance score is used to train a neural network as a surrogate model. To check, if this surrogate model has a high accuracy with respect to the original classifier, we also train neural networks with eight features randomly selected from the leftover features. Accuracy is defined as the percentage of correctly predicted cluster labels. A short introduction to neural networks and details on the architecture of the used neural network are given in the appendix \ref{sc:App_NN}. Training, test and validation data each contained $\frac{1}{3}$ of the full data. The networks were trained on (a) the eight optimal mode-mode features, (b) the eight optimal median features and (c) eight randomly selected features. For each group 100 networks were trained each of which was initialized with a different random seed to marginally change the performances in the groups (a) and (b). Also, each network in group (c) has a different selection of random features. This results in a distribution of network accuracies for each of the three groups. The eight correlations used for the mode-mode method are: E/M, E/F, E/HC, M/U, F/IT, CS/IT, U/T and U/IT (cf. the caption of figure \ref{fig:RelvantCorrelations} for the abbreviations). \\
The accuracy of predicting the original clustering label out of the eight clusters was chosen as a goodness criterion for the different feature selections. The results are depicted in the kernel density plots in figure \ref{fig:NN_comparison}, which are smoothened histograms of the 100 accuracy values for each neural network. The random feature selection achieves noticeably worse results than the two XAI based selection methods. While the mode-mode ($\mu_{mode} \pm \sigma_{mode}= 0.895\pm 0.004$) is slightly better than the median method ($\mu_{median}\pm \sigma_{median} = 0.882 \pm 0.005$), their confidence intervals (CIs) overlap at the $2\sigma$ level. Therefore, one cannot safely conclude that one of them is better than the other. However, their CIs for one $\sigma$ do not overlap and the mode-mode criterion possesses superior explainability for the individual clusters as demonstrated in figure \ref{fig:ElbowPlots_Mean_ModeMode}. Thus, it is recommended to use the mode-mode instead of the median.

\section{Discussion}
\label{sec:Discussion}

The goal of the research presented in this article was to answer the question whether it is possible to identify sector correlations that dominate the assignment of a trading day to a market state. To this end we used a clustering procedure that allows for an interpretation by a method from the field of XAI. We designed an aggregation procedure to combine these local explanations to global feature relevances for the entire data. The resulting relevance curves show pronounced elbow plots for the mode-mode aggregation like in figure \ref{fig:ElbowPlots_Mean_ModeMode} which can be analysed with a change point identification method to define cutoffs between irrelevant and relevant features.

Using the described methodology it is possible to show that the assignment of a certain trading day of the S\&P 500 market to a cluster (i.e. to a market state) is dominated by only a few sector correlations. The correlations involving the Energy or IT sector each make up three of the eight best features selected by the mode-mode XAI aggregation, respectively, but E/IT is not one of the eight best mode-mode features. Note that in figure \ref{fig:RelvantCorrelations}, clusters 1, 2, 5 and 6 also show a high influence of IT and/or Energy. The frequent occurrence of the IT sector among the highly relevant correlations may be traced back to the fact that the observation period includes the build-up, the emergence and burst of the dot-com bubble which saw the rise and rapid decline of many IT companies. This is also reflected in the analyses in \cite{Mnnix2012} and \cite{Stepanov2015}. The high importance of the energy sector implies that this sector has major influence in determining whether the economy is in a state of crisis or growth, probably because all other economic activity requires energy for e.g. transportation or production. While the high importance of the IT sector might have to be attributed to the dot-com bubble, the continued influence of the energy sector on the economic state might be a more reasonable conclusion from this analysis.
These results support previous studies that have identified the energy sector as a key influence in economic activity \cite{EnergyGrowthOECD,EnergyPoland,TaiwanEnergy}. \\
We also showed that a neural network which is only trained on the features with the highest XAI relevance scores can predict the trading day's cluster with a high degree of accuracy. In general our results imply that a further dimensionality reduction for the description of the dynamics of the financial market is possible. This can be particularly useful in high frequency trading applications, where the time span of interest lies within the life expectancy of a cluster. In this case, when data needs to be processed very quickly, it may be too slow to work with the complete correlation matrix. Nevertheless, our findings suggest that a reduced matrix only containing the most informative correlations for this cluster can lower the computation times without sacrificing too much accuracy. Note that the quantification of each feature's relevance was only possible due to the nature of the used XAI algorithm. The other discussed approaches could perform a dimensionality reduction only under the significantly higher effort of a less objective, since qualitative attempt which can lead to ambiguous results.

\subsection{Comparison to other Clustering Algorithms}

Our analysis follows the work in \cite{Mnnix2012,Rinn2015DynamicsOQ,Stepanov2015} and treats the correlation matrix of the returns as the observable that is to be sorted into clusters (market states) via k-means. A different approach to the identification of market states is given in \cite{Marsili_FinancialStatesViaReturns}, where the clustering algorithm in the maximum-likelihood-framework of \cite{Marsili_alg} is used. The author of \cite{Marsili_FinancialStatesViaReturns} not only uses the clustering algorithm to identify market states, but also to identify economic sectors. These show strong overlap with the sectors in the SIC classification which have been used as industry standards and recently been replaced by the GICS sectors. Because of this strong overlap, we use the GICS classification as a starting point for our analysis. In \cite{Marsili_FinancialStatesViaReturns}, daily prices from 1990 to 1999 have been analysed, whereas our work uses data from 1992 to 2012 in line with \cite{Rinn2015DynamicsOQ,Stepanov2015} and therefore has more recent data. While \cite{Marsili_FinancialStatesViaReturns} gauges the importance of business sectors for the market states by showing selected scatter plots of the companies' average returns for two different market states, we use an all-vs.-all comparison of all market states with the methodology of XAI. Despite the increased time frame and the different approach to clustering, our analysis confirms some of the key findings in \cite{Marsili_FinancialStatesViaReturns}, namely the pronounced role of the High-Tech companies (roughly corresponding to the IT sector in our analysis) and the sectors of Oil \& Gas and Petroleum companies (corresponding to the Energy sector). While \cite{Marsili_FinancialStatesViaReturns} also finds a peculiar behaviour of the Gold \& Silver mining companies, our analysis does not have such a high resolution of detailed business sectors and instead those companies are part of the Materials sector. While \cite{Marsili_FinancialStatesViaReturns} does not find any peculiar behaviour for the Finance sector (or Commercial Banks in \cite{Marsili_FinancialStatesViaReturns}), the Finance sector appears twice among our top 8 correlations, indicating an above average importance. This probably corresponds to the fact that our data, unlike the time interval used in \cite{Marsili_FinancialStatesViaReturns}, includes the 2008 financial crisis. \\
An interesting difference between our study and the analysis of market states in \cite{Mnnix2012} can be found by regarding the last state (8 in \cite{Mnnix2012} and 7 in our article): the qualitative state comparison in \cite{Mnnix2012} simply finds an overall high degree of correlation across the different sectors during this state, whereas our analysis still identifies that the two sector correlations Energy/Finance and Utilities/IT (cf. figure \ref{fig:Compare_Centroids_XAI}) have a dominant effect on this cluster. The identification of those two dominant correlations was only possible due to the quantitative approach of the XAI methodology.

\subsection{Outlook}

While this article used a standard k-means clustering algorithm and therefore deviated from the hierarchical clustering in  \cite{Mnnix2012, Rinn2015DynamicsOQ,Stepanov2015}, this opens up another possibility for follow-up research by combining these methods: using a hierarchical clustering and using the XAI on each split in the hierarchy may increase our understanding even further via XAI as an addition to the intrinsic interpretability of hierarchical algorithms. This would potentially tie together the explainability of XAI with the paradigm of interpretable models presented in \cite{Rudin2019}.\\
Despite the insightful analyses of financial correlation matrices in \cite{Mnnix2012, Rinn2015DynamicsOQ,Stepanov2015} and in this article, one has to keep in mind that the analysis of the states is based on the matrix of Pearson's correlation coefficients between financial time series. Although this method is frequently used in financial research \cite{Markowitz1952, RandomMatrixLalouxPotters, RandomMatrixStanley, Review_CorrelationsHierarchiesClustering_FinancialMarkets}, it has been subject to criticism: As noted in \cite{SaoPauloStock}, Pearson's correlation only provides inference on linear correlation, but not on nonlinear correlations. This can also be seen in the stylised fact that returns are uncorrelated, but squared returns are strongly correlated \cite{MantegnaStanleyBook}. Instead, Spearman's correlation may provide another view on correlations and include nonlinear effects \cite{SANDOVAL2012187} and therefore can be used as another starting point for follow-up research. Additionally, as already mentioned in section \ref{sec:DataPrep}, the GICS classification has been updated over time to reflect changes in the economy's structure. Because these updates also happened during the regarded 20 years time period, they may have affected the results of this analysis. Moreover, there is a more subtle problem about the data that can be referred to as a survivorship bias: Like \cite{Mnnix2012, Rinn2015DynamicsOQ,Stepanov2015}, we only regard companies whose stock prices are available for the entire time period. Companies that went bankrupt or were merged into another company during the time period are therefore completely disregarded by these analyses, although their bankruptcies or takeovers certainly contain valuable information about the state of the economy.

\section*{Acknowledgements}
Tobias Wand thanks Shanshan Wang, Henrik Bette (both University of Duisburg-Essen) and Katrin Schmietendorf (WWU Münster) and the anonymous referees for their valuable suggestions. Tobias Wand and Martin Heßler are funded by the Studienstiftung des deutschen Volkes.

\FloatBarrier

\begingroup
\setlength{\emergencystretch}{12em}
\printbibliography

\newpage
\FloatBarrier
\appendix

\section{Appendix: Details on Layer-Wise Relevance Propagation for K-Means}
\label{sc:APP_LRP}

These sections give a more detailed explanation of how to reformulate the k-means clustering algorithm and how to use the Layer-wise Relevance Propagation as described in \cite{Kauffmann2022}. 

\subsection{Neuralisation of K-Means}
It has been proven in the supplementary materials of \cite{Kauffmann2022} that the k-means decision criterion \eqref{eq:KMeansInequality} of cluster $j$ vs. the other clusters $l\neq j$ can be rewritten in two layers and a third decision layer. This procedure is described by algorithm \ref{alg:NeuralisedKMeans}. Note that while the k-means classifier tests all clusters against each other, the neuralised classifier only tests if an instance belongs to a specific cluster $j$ versus all other clusters $l\neq j$. If the instance does not belong to the $j^\textmd{th}$ cluster, algorithm \ref{alg:NeuralisedKMeans} does not say which of the other clusters is correct. Therefore, one needs as many neuralised classifiers as there are clusters to fully neuralise equation \eqref{eq:KMeansInequality}.

\begin{algorithm}
  \caption{Neuralisation of a K-Means Classifier}\label{alg:NeuralisedKMeans}

  \begin{algorithmic}[1]
\Statex \textbf{Goal:} Test assignment of \textbf{C} to cluster $j$ against all alternative clusters $l\neq j$.
\Statex \textbf{Input:} Instance \textbf{C} and cluster centroids $(\textbf{C}_l)_l$
\Statex \textbf{Procedure in three layers:}
    \State Calculate $\forall l\neq j$ $h_l = \textbf{w}_l \cdot \textbf{C} + b_l$  with $\textbf{w}_l = 2(\textbf{C}_j - \textbf{C}_l )$ and $b_l = ||\textbf{C}_l||^2 -||\textbf{C}_j||^2 $
    \State Calculate $f_j =  \min_{l\neq j} h_l$ 
    \State assign $\textbf{C}$ to cluster $j$, if $f_j > 0$  
  \end{algorithmic}
\end{algorithm}

\subsection{Layer-wise Relevance Propagation}
Layer-wise Relevance Propagation is an XAI method that runs backwards through a neural network. At each step, it calculates the relevance that the current neuron has for the deeper layer and distributes it accordingly to the next higher layer's neurons. A conservation property ensures that the combined relevance is preserved while being propagated to the higher layers and is illustrated in figure \ref{fig:LRP_sketch}. According to \cite{LRP_book}, this reasoning is analogous to Kirchhoff's laws in electrical circuits and used similarly in other XAI techniques such as \cite{HierarchicalNetworkInterpretation, PropagatingActivationDifferences,TopDownNeuralAttention}.\\
For the neuralised k-means algorithm \ref{alg:NeuralisedKMeans}, \cite{Kauffmann2022} suggests the following propagation rule of relevance $\rho_l$ from $f_j$ backwards to $h_l$:

\begin{equation}
    \label{eq:PropagationRule1}
    \rho_l = \frac{\exp(-\beta h_l)}{\sum_{l\neq j} \exp(-\beta h_l)} f_j 
\end{equation}
with the inverse mean $\beta = \mathbb{E}[f_j]^{-1}$ of $f_j$ over the entire data. The second rule for the propagation of relevance $\rho_l$ of each node $h_l$ to the original input features $c_i$ of $\textbf{C}$ is given by

\begin{equation}
    \label{eq:PropagationRule2}
    \rho_i = \sum_{l\neq j} \frac{(c_i - m_{i,l})w_{i,l}}{\sum_i(x_i - m_{i,l})w_{i,l}} \rho_l
\end{equation}
with $\textbf{m}_l = (\textbf{C}_j + \textbf{C}_l)/2$ as the point directly between the centroids and $w_{i,l}$ as the $i^\text{th}$ component of the difference vector $\textbf{w}_l$ between the cluster centroids $j$ and $l$ (cf. the pseudocode in \ref{alg:NeuralisedKMeans}). Note that in our study, each $c_i$ is a correlation like in equation \eqref{eq:Correlation}.

\begin{figure}[h!]
    \centering
    \includegraphics[width = 0.7\textwidth]{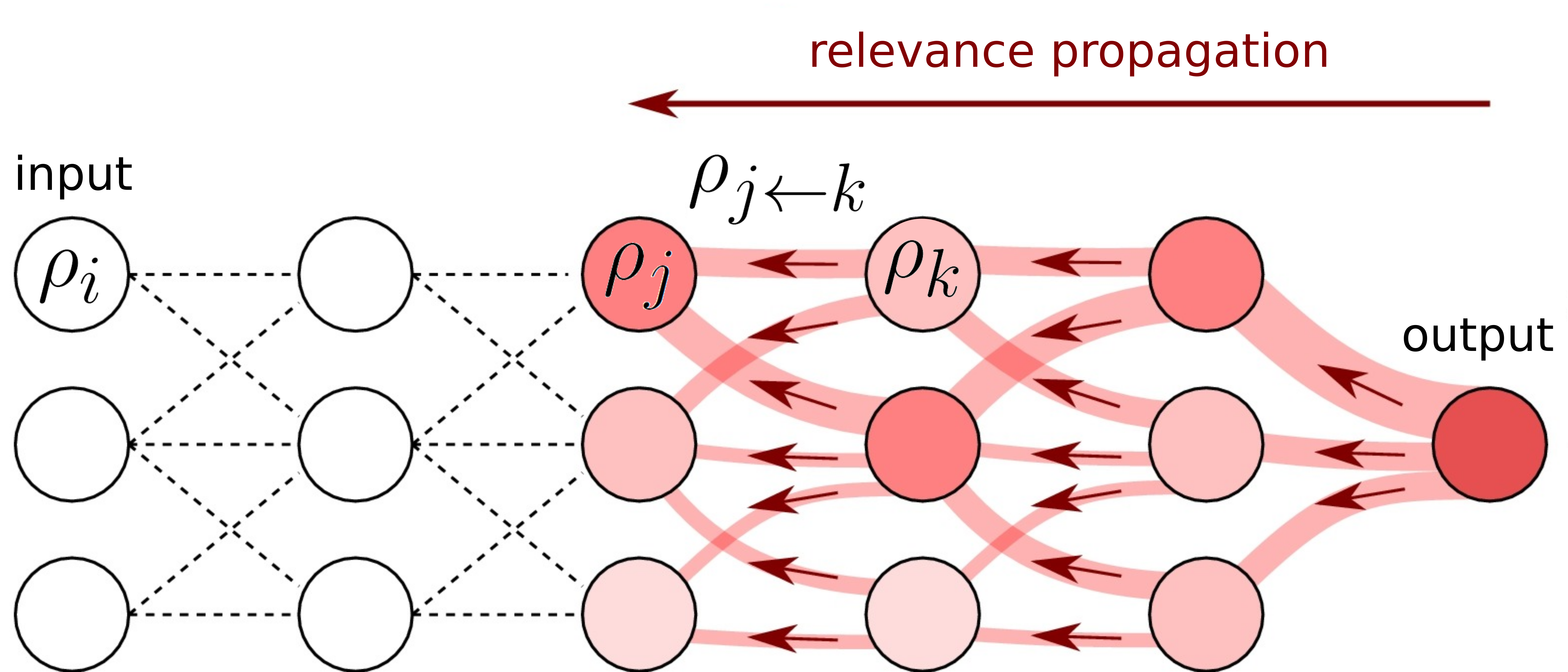}
    \caption{Illustration of the Layer-wise Relevance Propagation, taken from \cite{MONTAVON20181} and adapted under a CC BY 4.0 license (\url{https://creativecommons.org/licenses/by/4.0/}) to stay consistent with our article's notation.}
    \label{fig:LRP_sketch}
\end{figure}

\section{Appendix: Elbow Plots}
\label{sc:App}

The same plots as in figure \ref{fig:ElbowPlots_Mean_ModeMode} are depicted in this section for the other seven clusters. While a pronounced elbow is visible for all clusters in the mode-mode aggregation, the median method often shows a continuum.

\begin{figure}[H]
    \centering
    \includegraphics[width = 0.9\textwidth]{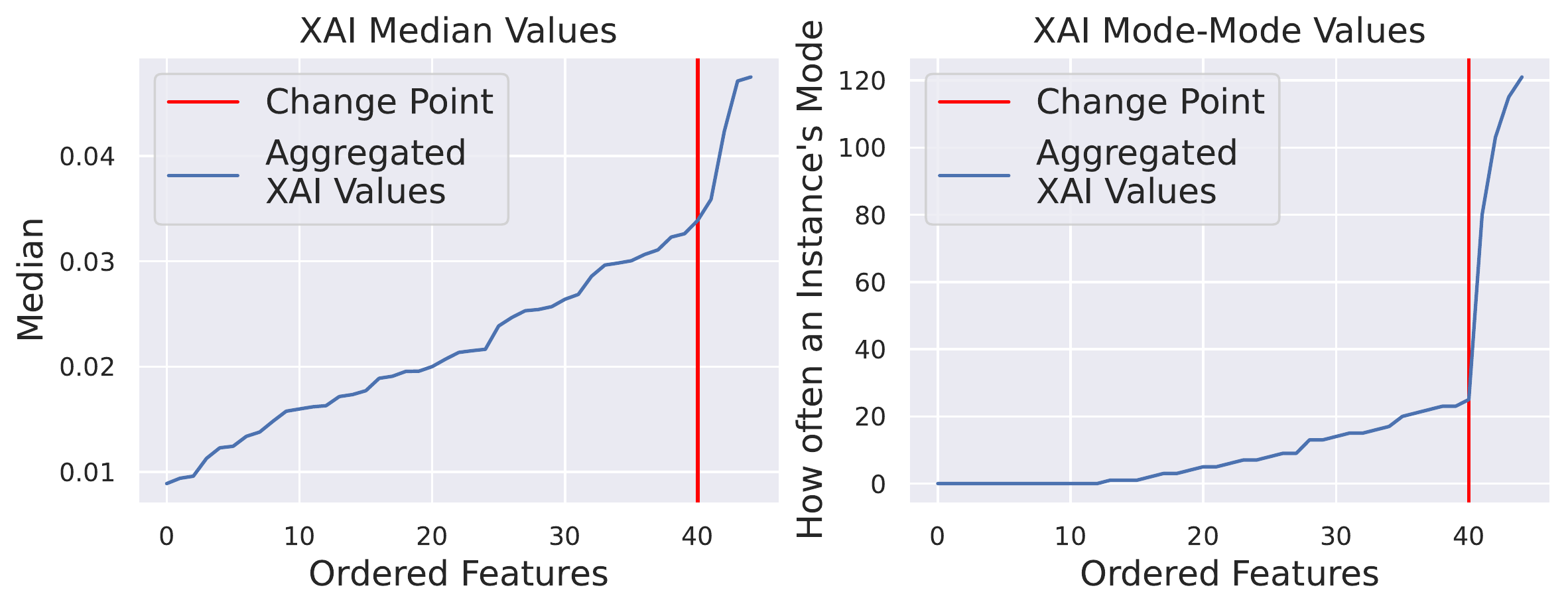}
    \caption{The same figures as in figure \ref{fig:ElbowPlots_Mean_ModeMode}, but now for cluster 0.}
    \label{fig:Appendix0}
\end{figure}

\begin{figure}[H]
    \centering
    \includegraphics[width = 0.9\textwidth ]{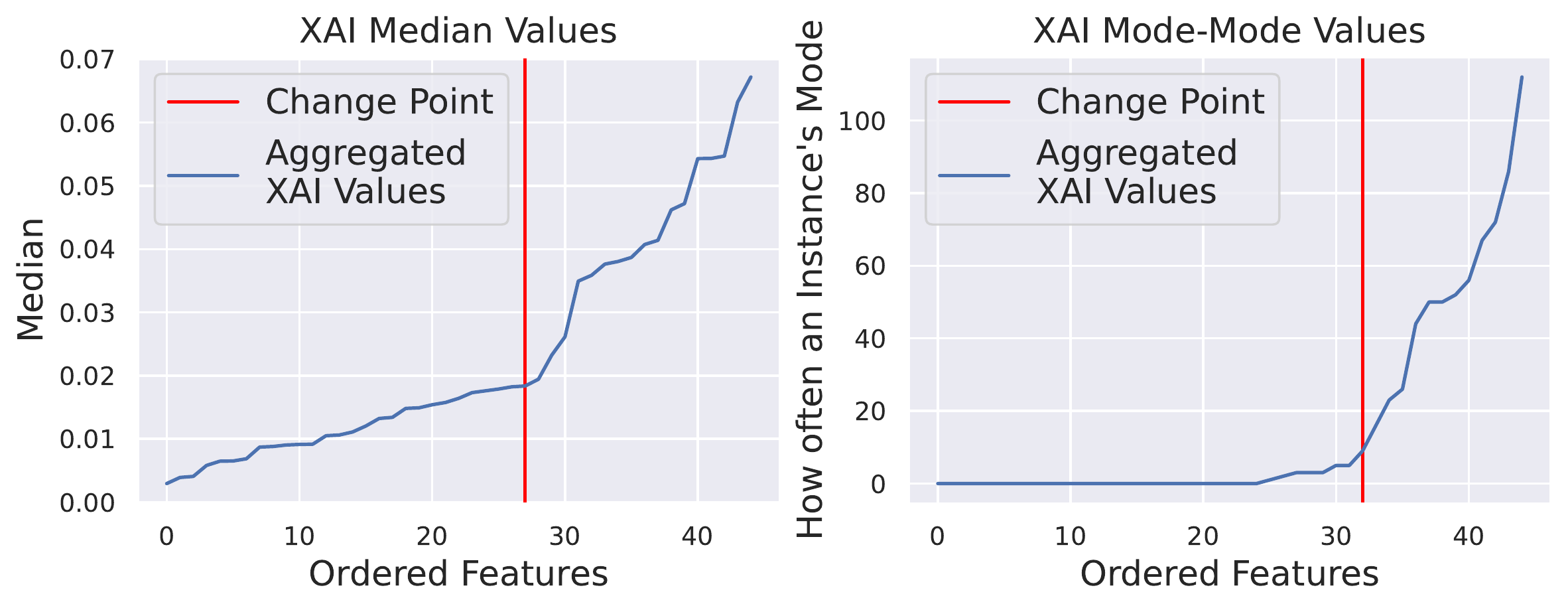}
    \caption{The same figures as in figure \ref{fig:ElbowPlots_Mean_ModeMode}, but now for cluster 1.}
    \label{fig:Appendix1}
\end{figure}

\begin{figure}[H]
    \centering
    \includegraphics[width = 0.9\textwidth]{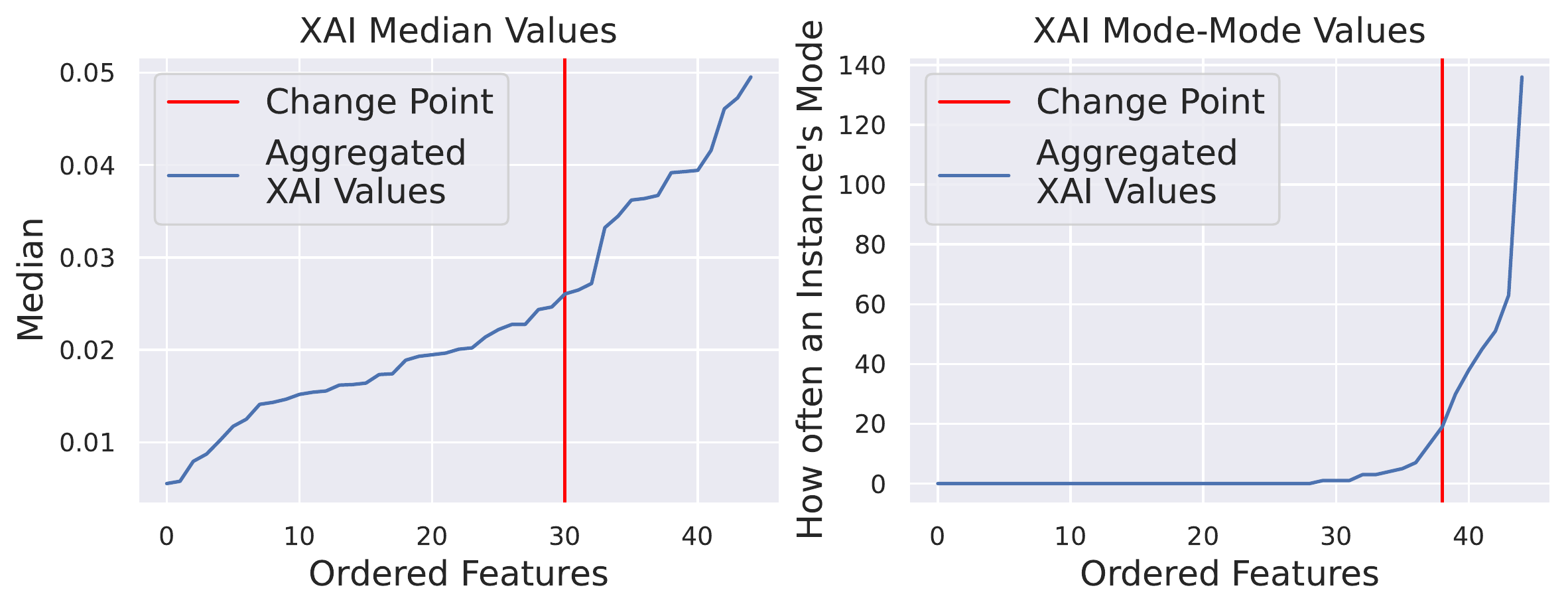}
    \caption{The same figures as in figure \ref{fig:ElbowPlots_Mean_ModeMode}, but now for cluster 3.}
    \label{fig:Appendix3}
\end{figure}

\begin{figure}[H]
    \centering
    \includegraphics[width = 0.9\textwidth]{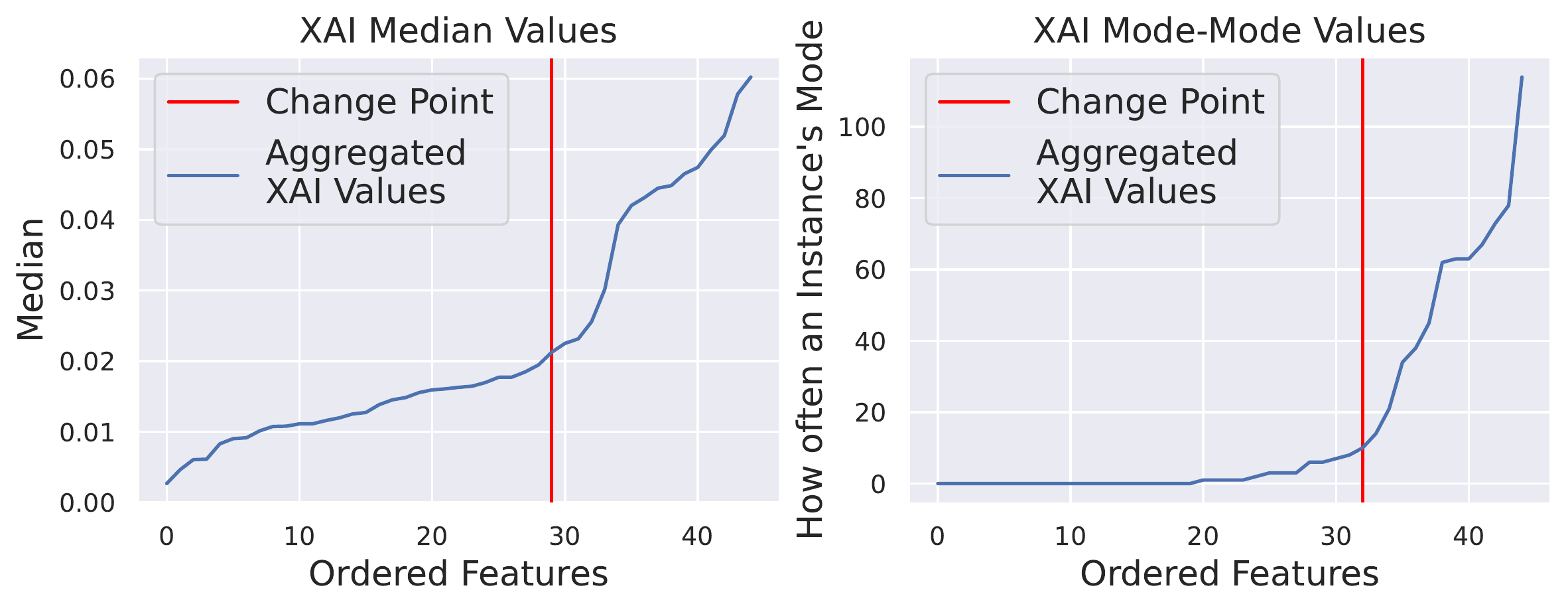}
    \caption{The same figures as in figure \ref{fig:ElbowPlots_Mean_ModeMode}, but now for cluster 4.}
    \label{fig:Appendix4}
\end{figure}

\begin{figure}[H]
    \centering
    \includegraphics[width = 0.9\textwidth]{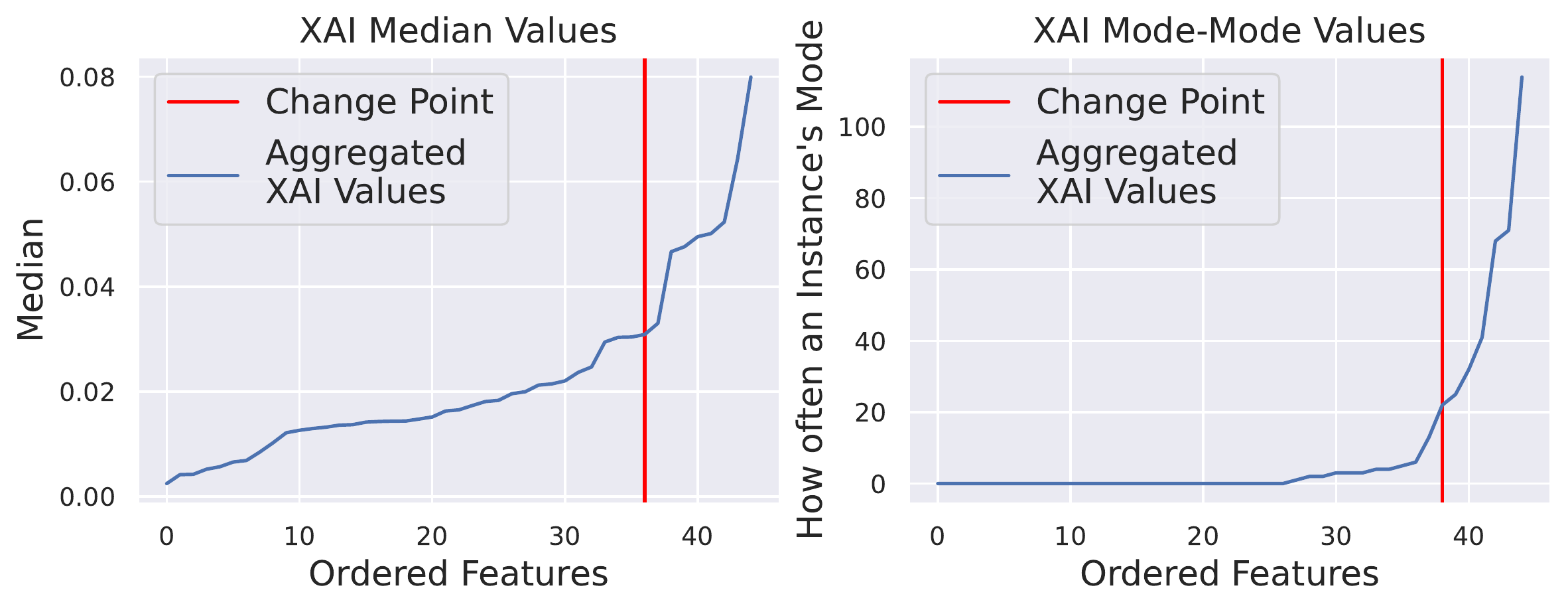}
    \caption{The same figures as in figure \ref{fig:ElbowPlots_Mean_ModeMode}, but now for cluster 5.}
    \label{fig:Appendix5}
\end{figure}

\begin{figure}[H]
    \centering
    \includegraphics[width = 0.9\textwidth]{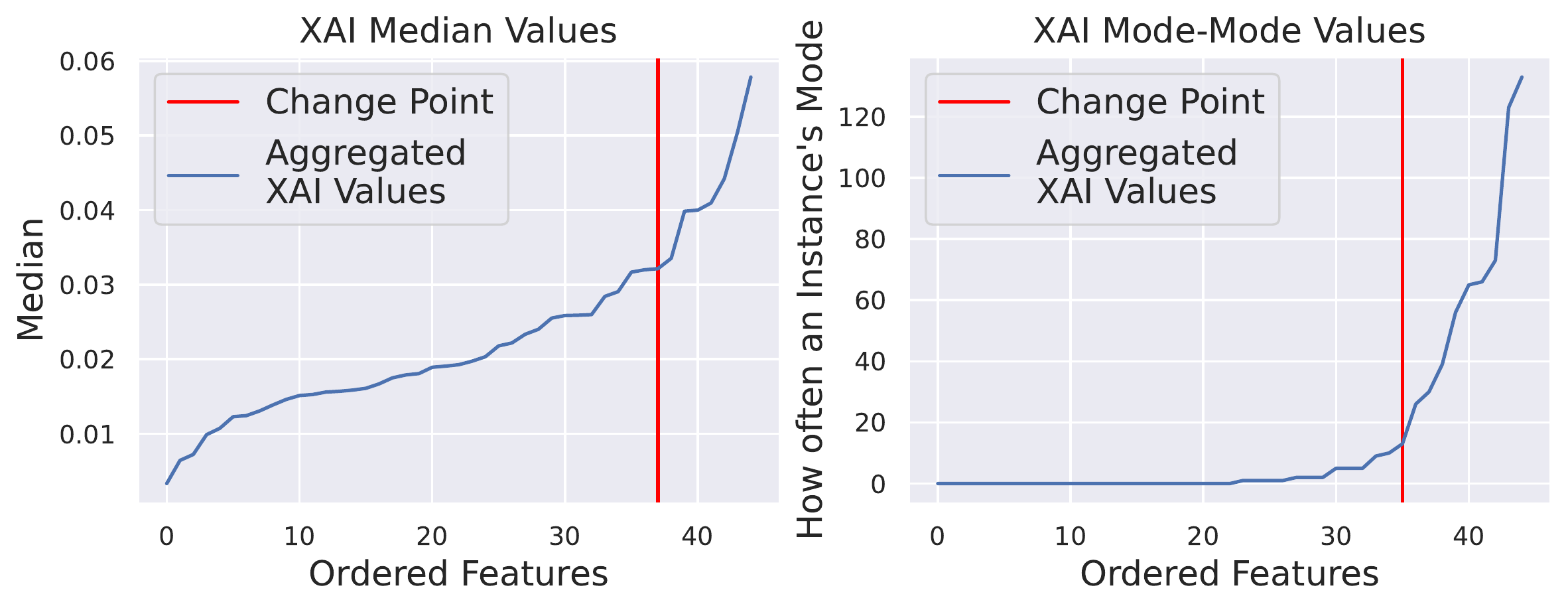}
    \caption{The same figures as in figure \ref{fig:ElbowPlots_Mean_ModeMode}, but now for cluster 6.}
    \label{fig:Appendix6}
\end{figure}

\begin{figure}[H]
    \centering
    \includegraphics[width = 0.9\textwidth]{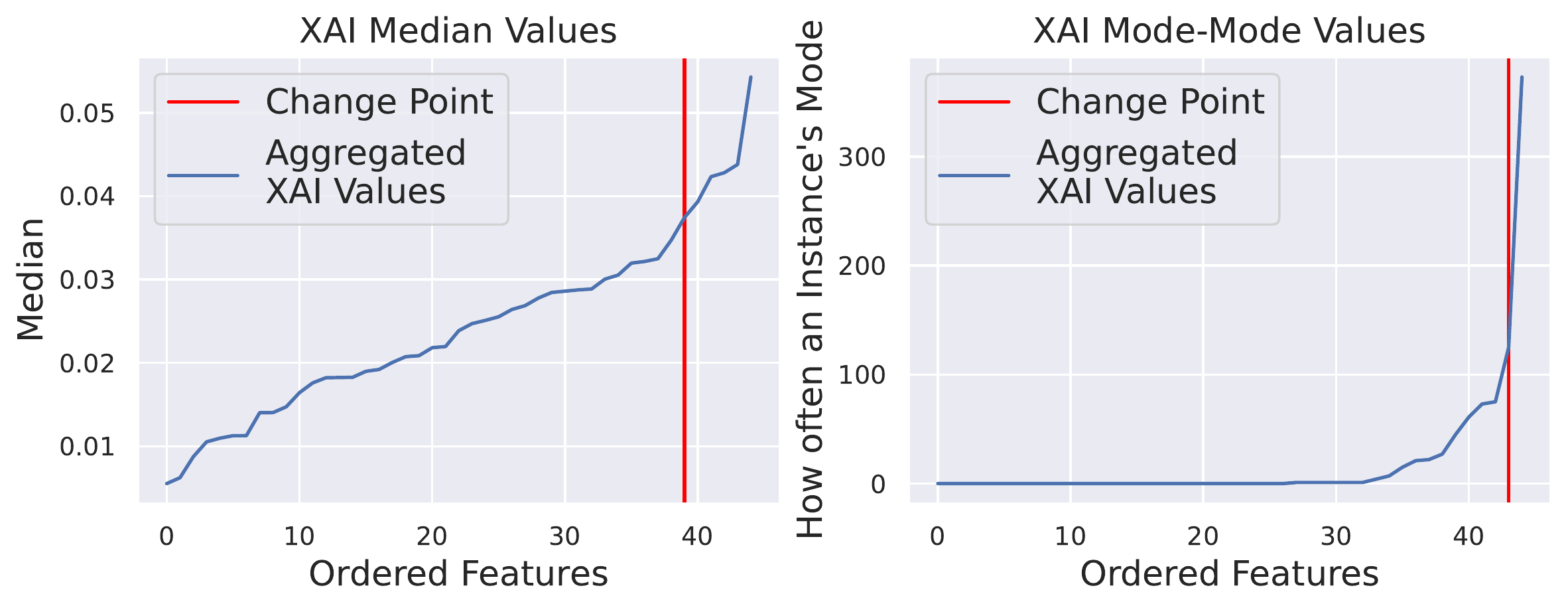}
    \caption{The same figures as in figure \ref{fig:ElbowPlots_Mean_ModeMode}, but now for cluster 7.}
    \label{fig:Appendix7}
\end{figure}

\section{Appendix: Neural Network Architecture}
\label{sc:App_NN}

The neural network used as a surrogate model in this article is a deep multi-layer perceptron network for a classification task to predict which of the eight clusters the instance belongs to. The architecture of such a network is as follows: The basic unit of the network is a neuron or perceptron, which gets an input $z\in\mathbb{R}$ and produces an output $u \in\mathbb{R}$ via an activation function $f(z) = u$. Motivated by biological neurons, the activation functions often only produce a significantly nonzero output (the neuron "fires"), if the input is larger than some threshold. Common choices for activation functions are given by

\begin{align}
\label{eg:Activations}
    &\text{relu}(z) = \max(0,z) \hspace{10mm}
    \text{LeakyReLU}_\alpha(z) = \max(\alpha z, z) \textmd{ with free parameter }\alpha \\ \nonumber &\text{selu}_{(s,a)}(z) = sz \mathbbm{1}_{z\leq0} + sa(\exp{(z)}-1)\mathbbm{1}_{z<0} \textmd{ with }  a=1.67326324 \textmd{ and } s=1.05070098\textmd{.}
\end{align}
The neurons are organised in layers where the output of the $N_n$ neurons in layer $n$ is used as inputs for the $N_{n+1}$ neurons in the $n+1^\text{th}$ layer. If the connections are \textit{dense}, then all $N_n$ previous outputs are combined in linear combinations to contribute to the next layer's inputs. The weights of the linear combinations are then optimised to achieve the best accuracy. If the connections are in a \textit{dropout} layer of rate $\beta$, then at every optimisation step, $\beta$ of the connections are forcibly set to $0$ to prevent an overfitting of the network (i.e. that the network is fit so closely to the known data that it does not generalise well on unknown data).\\
The very first layer (input layer) receives the features (observables) of the data and, in this case, the output layer is giving a prediction about which of the 8 clusters the data belongs to.\\ The softmax function is often used to get from the continuous values of the neurons' outputs to an actual prediction. If the network has to choose between $k$ possible predictions, then it will end with $k$ linear combinations $l_1,\dots,l_k$ of the neurons' outputs in its penultimate layer. For the $k$ possible classes it transforms these linear combinations into probabilities $p_1,\dots,p_k$ via the softmax function following
\begin{equation*}
    \textmd{softmax}(l_i) = \frac{\exp{(l_i)}}{\sum_i \exp{(l_i)}}.
\end{equation*}
Note that $\sum_i p_i=1$. Finally, it uses the highest probability as its prediction. Further details on the construction of neural networks can be found e.g. in \cite{Geron2019-gw} and technical details on the implementation of the neural network are given in \cite{chollet2015keras, tensorflow2015-whitepaper}. The network architecture used in this article is detailed in table \ref{tab:NN_Architecture}.

\begin{table}
    \centering
    \begin{tabular}{c|ccccccc}
        Layer & Dense & Dense & Dense & Dense & Dense & Dropout & Dense \\ \hline
        Units & 256 & 128 & 128 & 1024 & 128 & - & 8 \\
        Activation & selu & relu & relu & LeakyReLu & LeakyReLu & - & softmax \\
        Parameter & - & - & - & 0.05 & 0.01 & 0.3 & - 
    \end{tabular}
    \caption{Architecture of the neural network with 100 training epochs for the surrogate model in section \ref{sec:Results_XAI} from left to right. There are eight input units for the eight features of the surrogate model and eight output units for the eight clusters.}
    \label{tab:NN_Architecture}
\end{table}

\end{document}